\def\diy{\displaystyle}

\def\C{\mathbb C}
\def\D{\mathbb D}
\def\E{\mathbb E}

\def\L{\mathbb L}

\def\N{\mathbb N}
\def\P{\mathbb P}
\def\R{\mathbb R}

\def\Z{\mathbb Z}

\def\ovln{\overline}

\def\wh{\widehat}

\def\cL{{\mathcal L}}

\def\cN{{\mathcal N}}

\def\cS{{\mathcal S}}

\def\mfM{{\mathfrak M}}

\def\dist{{\rm{dist}}}

\def\one{{\mathbf 1}}

\def\oj{\overline j}
\def\oU{\overline U}
\def\oV{\overline V}

\def\oW{\overline W}
\def\u0{{\underline 0}}

\def\uu{{\underline u}}

\def\ux{{\underline x}}

\def\omu{{\overline\mu}}

\def\wh{\widehat}

\def\eps{{\epsilon}}
\def\om{{\omega}}
\def\Lam{{\Lambda}}
\def\Gam{{\Gamma}}

\def\bfv{\mathbf v}

\def\rA{\rm A}
\def\rb{\rm b}

\def\ru{\rm u}

\def\rx{\rm x}
\def\ry{\rm y}

\def\pmn{\par\medskip\noindent}
\def\psn{\par\smallskip\noindent}
\def\z2{{\Z^2}}
\def\zp2{{\Z^2_{\geq}}}

\def\esm#1{{\E\left[ \, #1 \, \right]}}

\def\dist{{\,{\rm dist}}}

\def\bfv{\mathbf v}
\def\bfV{\mathbf V}

\def\BGam{{\mbox{\boldmath${\Gamma}$}}}
\def\Bgam{{\mbox{\boldmath${\gamma}$}}}
\def\BLam{{\mbox{\boldmath${\Lam}$}}}
\def\Bphi{{\mbox{\boldmath${\phi}$}}}
\def\Bpsi{{\mbox{\boldmath${\psi}$}}}

\def\RFCA{{random field}}
\def\RFCAs{{random fields}}

\documentclass{birkmult}

\usepackage{color}
\definecolor{gr}{gray}{0.50}
 \newtheorem{thm}{Theorem}[section]
 
 \newtheorem{lem}{Lemma}[section]

 \theoremstyle{definition}
 
 \theoremstyle{remark}

 \numberwithin{equation}{section}

\begin{document}
%
%
%
%
%
%
%
%
%
\title[Wegner-type bounds for a continuous Anderson model]
 {Wegner-type bounds for a two-particle\\
Anderson model in a continuous space}
\author[A. Boutet de Monvel]{A. Boutet de Monvel}

\address{%
Institut de Math\'{e}matiques de Jussieu\\
Universit\'{e} Paris Diderot Paris 7\\
175 rue du Chevaleret, 75013 Paris, France}

\email{aboutet@math.jussieu.fr}

\author{V. Chulaevsky}
\address{D\'{e}partement de Math\'{e}matiques et Informatique, \\
Universit\'{e} de Reims, Moulin de la Housse, B.P. 1039, \\
51687 Reims Cedex 2, France }
\email{victor.tchoulaevski@univ-reims.fr}

\author{Y. Suhov}
\address{Statistical Laboratory, DPMMS\\
University of Cambridge, Wilberforce Road, \\
Cambidge CB3 0WB, UK}
\email{Y.M.Suhov@statslab.cam.ac.uk}



\date{January 1, 2004}

\begin{abstract}
We analyse a two-particle quantum system in $\R^d$ with
interaction and in presence of a random external potential field
with a continuous argument (an Anderson model in a continuous space).
Our aim is to establish the so-called Wegner-type estimates
for such a model, assessing the probability that random spectra of
Hamiltonians in finite volumes intersect with a given set.
For the lattice version of the two-particle model, a similar
result was obtained in \cite{CS1}

\end{abstract}

\maketitle

\section{Introduction. The two-particle Anderson\\ Hamiltonian
in a continuous space}
\label{intro}

This paper is a follow-up of \cite{CS1}
and establishes Wegner-type (more precisely, Wegmer--Stollmann-type)
bounds for random continuous Schr\"{o}dinger operators.
We focus here on
a two-particle interactive Anderson model in a continuous space,
subject to a random external field with a continuous argument.
The infinite-volume Hamiltonian of the model
is a Schr\"{o}dinger operator $H\left(=H^{(2)}(\om )\right)$
acting on functions $\Bphi\in\L_2(\R^d\times\R^d)$:
$$H\Bphi (\ux)=H^0\Bphi (\ux)+
W(\ux ;\om )\Bphi (\ux ),\;\;\ux=(x_1,x_2)\in\R^d\times\R^d.
\eqno (1.1)$$
Here $H^0$ is the kinetic energy operator:
$$H^0=-\diy{\frac{1}{2}}\sum\limits_{j=1,2}\Delta_j,
\eqno (1.2)$$
where $\Delta_j$ is the Laplacian in variable $x_j
=\big({\rx}_j^{(1)},\ldots,{\rx}_j^{(d)}\big)\in\R^d$ corresponding
to the $j$th particle:
$$\Delta_j=\sum_{i=1}^d\diy{\frac{\partial^2}{\partial
{{\rx}_j^{(i)}}^2}}\,,\eqno (1.3)$$

The randomness in Hamiltonian $H^{(2)}(\om )$ is concentrated in the
potential energy function $W(\ux;\om )$ which is written in
the form
$$W(\ux ;\om )=U(\ux)+\sum_{j=1,2} V(x_j;\om),
\;\;\ux =(x_1,x_2)\in\R^d\times\R^d,\eqno(1.4)$$
The term $U(\ux )$ represents the (nonrandom) interaction potential
while the sum $\big[V(x_1;\om)+V(x_2;\om )\big]$ describes
the action of an external potential field on the two-particle system.
More precisely, the external potential field is represented by a
family of real random variables (RVs) $\bfV=\{V(x;\om ),x\in\R^d\}$, or, in
probabilistic terminology, by a real-valued random field with
a continuous argument (or simply \RFCA). A sample of such a random field is
a (measurable) function $\bfv:\;x\in\R^d\mapsto\R$, and its its graph is
a `hypersurface' in $\R^d\times\R$, with a unique point of
intersection with any straight line orthogonal to the `argument
space' $\R^d$. The argument $\om$ in the notation $V(x;\om )$
stresses randomness of the external potential. The distribution
of the \RFCA $\,\bfV$ is denoted by $\P$: it is a probability measure
on a (suitably chosen) sigma-algebra $\mfM$ in the space of
sample functions $\bfv:\;R^d\to\R$. Formal definitions are
given in Section \ref{SectExtPot}; cf. \cite{AT}. A classical example
is a Gaussian distribution on $\mfM$; see, e.g., \cite{Adl}.

The randomness introduced in Hamiltonian $H$ perplexes its spectral
properties. A physically-motivated result expected here is
that the spectrum of $H$ near its `lower edge' is
pure point with probability one, and the corresponding eigenfunctions
decay exponentially in space. (This is called Lifshits-tail-based
exponential localisation.) Similar results have been established
for a variety of single-particle models where the Hamiltonian
acts on a function $\phi\in\L_2(\R^d)$ as
$-\Delta\phi (x)/2 +V(x;\om )\phi (x)$, $x\in\R^d$. See \cite{St2}, \cite{LMW} and references therein.
As far as localisation is concerned, the main difference between
a single-particle model and its two-particle counterpart is that
replacing term $V(x;\om )$ by the sum $W(\ux;\om )=U(\ux )+\sum\limits_{j=1,2}
V(x_j;\om )$ leads to `strong' dependencies between RVs
$W(\ux;\om )$ and $W(\ux';\om )$, no matter how far points
$\ux,\ux'\,\in\R^d\times\R^d$ are positioned from each other.
The ensuing difficulty cannot be relieved no matter how
quickly correlations between RVs $V(x;\om )$ and $V(x';\om )$ decay
when $x,x'\in\R^d$ are far from each other. Cf. \cite{CS1, CS2} where this
discussion has been conducted for the so-called tight-binding
two-particle Anderson model on a lattice.

An important ingredient of the existing localisation proofs for
single-particle Anderson models or their two-particle counterparts
is the so-called Wegner-type (or Wegner--Stollmann type)
estimates for the corresponding
Hamiltonian. [Exceptions are one-dimensional models where
the whole physics of localisation is rather special.] This explains
the title of this paper. The
Wegner-type estimates are produced for the eigen-values of
a finite-volume approximation of the Hamiltonian; cf.
\cite{FHLM}. In our situation,
these are operators $H_\BLam\left(=H^{(2)}_{\BLam}(\om )\right)$
acting on functions $\Bphi\in\L_2(\BLam )$:
$$H_\BLam\Bphi (\ux)=H^0_\BLam\Bphi (\ux)+
W(\ux ;\om )\Bphi (\ux ),\;\;\ux=(x_1,x_2)\in\BLam ,\eqno (1.5)$$
with
$$H^0_\BLam =-\diy{\frac{1}{2}}\sum\limits_{j=1,2}\Delta^{(\BLam)}_j.
\eqno (1.6)$$
and $W(\ux;\om )$ defined, as before, by Equation (1.2.2).
Here $\BLam$ is a bounded domain in $\R^d\times\R^d$ which we choose
to be the Cartesian product of two $d$-dimensional cubes: for
$\uu =(u_1,u_2)\in\R^d\times\R^d$ and $L_1,L_2\in (0,\infty )$,
$$\BLam =\BLam_{L_1,L_2}(\uu )=\Lam_{L_1}(u_1)\times\Lam_{L_2}(u_2),
\eqno (1.7)$$
where, for $u=\big({\ru}^{(1)},\ldots ,{\ru}^{(d)}\big)\in\R^d$
and $L\in (0,\infty)$,
$$\Lam_L(u)={\operatornamewithlimits{\times}\limits_{i=1}^d}
\big[-L+{\ru}^{(i)},{\ru}^{(i)}+L\big].
\eqno (1.8)$$
The shorthand notation $\BLam$ for $\BLam_{L_1,L_2}(\uu )$ (and
$\BLam'$ for $\BLam_{L'_1,L'_2}(\uu')$) will be systematically used in this
paper. Further, $\Pi_1\BLam$ and
$\Pi_2\BLam$ will stand for cubes $\Lam_{L_1}(\uu_1)$ and
$\Lam_{L_2}(\uu_2)$ forming the projections of `two-particle' parallelepiped
$\BLam =\BLam_{L_1,L_2}(\uu )$ onto single-particle configurational spaces
(and similarly with $\Pi_1\BLam'$ and
$\Pi_2\BLam'$).

Next, $\Delta^{(\BLam )}_j$ is the Laplacian in variable $x_j
=\big({\rx}_j^{(1)},\ldots,{\rx}_j^{(d)}\big)\in\Pi\BLam$ confined
to cube $\Pi_j\BLam$, $j=1,2$, with a specified boundary condition
which we choose to be Dirichlet's:
$$\Delta^{\BLam}\left(=\Delta^{(\Pi_j\Lam )}_j\right)
=\sum_{i=1}^d\diy{\frac{\partial^2}{\partial
{{\rx}_j^{(i)}}^2}}+\hbox{Dirichlet's boundary condition.}\eqno (1.9)$$
In fact, a possible choice of boundary conditions is rather
broad and includes periodic and `elastic', in particular,
Neumann's. (What we need for our method to work is that the
self-adjoint extension specified by a boundary condition is
half-bounded and has a compact resolvent $(H_\BLam -zI)^{-1}$
for $z\in\C$ away from the real line.) See \cite{CL}, \cite{St2} for details.

Thus, operator $H^0_\BLam$ in Eqn (1.6) is understood in terms of the
sesquilinear form
$\left\langle\Bphi,H^0_\BLam\Bpsi\right\rangle$ in
$\L_2(\BLam )$, where $\langle\;\cdot\;,\;\cdot\;\rangle$ stand for
the standard inner product. A similar meaning is attributed to the
multiplication
operator by $W(\ux;\om )$ in $\L_2(\BLam )$; consequently,
we treat the function
$W(\ux;\om )$, $\ux\in\BLam$, as an element of $\L_2(\BLam )$
defined for $\P$-a.a. $\om$. Additional conditions imposed
below mean that this function is bounded on $\BLam$, and
the supremum $\oW_\BLam (\om )$ of its absolute value
has certain moments; see below.

Moreover, under the assumptions introduced in this paper,
the (self-adjoint) operator $H_\BLam$ is bounded from below
and has, with probability one, a discrete  spectrum
of a finite multiplicity. It is convenient to write its eigenvalues
$E^{({\BLam})}$ in an increasing order:
$$E^{({\BLam})}_0\leq E^{({\BLam})}_1\leq E^{({\BLam})}_2\leq\ldots .
\eqno (1.10)$$
In the `one-volume' Wegner estimate one
assesses the probability that at least one eigenvalue
$E^{(\BLam )}_k$ of operator $H_\BLam$ falls in
a (narrow) interval around a given point $E$ on the
spectral axis:
$$\P\left(\exists\;\hbox{ $k$
with $\left|E-E^{(\BLam )}_k\right|$}\leq\eps\right),\eqno (1.11)$$

Next, the `two-volume' Wegner estimate addresses the
probability  that the eigenvalues $E^{(\BLam )}_k$
and $E^{(\BLam')}_l$ of operators  $H_{\BLam}$
and $H_{\BLam'}$ come near to each other in a
given interval $J\subseteq\R$, for two (distant)
parallelepipeds $\BLam =\BLam_{L_1,L_2}(\uu)$ and
$\BLam'=\BLam_{L'_1,L'_2}(\uu ')$. That is,
$$\P\left(\exists\;\hbox{ $k$ and $l$
with $\E^{(\BLam )}_k,E^{(\BLam')}_l\in J$ and
$\left|E^{(\BLam )}_k-E^{(\BLam')}_l\right|$}\leq\eps\right),
\eqno (1.12)$$
Here and below,
$\P$ stands for the corresponding probability measure on the
underlying probability space (see Section \ref{SectExtPot}).

 From the probabilistic point of view, the estimates for
probabilities (1.11) and (1.12) are examples of concentration
inequalities, albeit for rather implicit RVs $E^{(\BLam )}_k$
carrying a considerable amount of dependence. For single-particle
Anderson models, assuming natural properties
of the random term $V(x;\om )$,
the Wegner estimates are rather straightforward.
For the two-particle models, the study of these estimates,
with a view of localisation, started in \cite{K}, \cite{CS1}, \cite{BCSS1} and is
continued in this paper.

In the main body of this work, the interaction potential $U$
satisfies the following property:
\pmn

(I) {\sl $U$ is a (measurable) bounded real function
$\R^d\times\R^d\to\R$ obeying}
$$
U(\ux )=U(\sigma\ux),\;\;\ux\in\R^d\times\R^d,\;\;
U(\ux)=0,\;\hbox{ if }\;\|x_1-x_2\|_{\max}>r_1.
\eqno (1.13)
$$
Here $r_1\in (0,\infty )$ is the interaction radius,
$\sigma$ stands for the permutation of the (vector) variables
($\sigma\ux =(x_2,x_1)$ for $\ux=(x_1,x_2)$) and $\|x_1-x_2\|_{\max}$
denotes the max-norm in $\R^d$: for $x_j=\left({\rx}_j^{(1)},\ldots ,
{\rx}_j^{(d)}\right)\in\R^d$, $j=1,2$:
$$
\|x_1-x_2\|_{\max}=\max\;\left[\left|{\rx}_1^{(k)}
-{\rx}_2^{(k)}\right|:\;k=1,\ldots ,d\right]
\eqno (1.14)
$$

We can also allow the case where $U$ has a hard core, i.e.,
$$U(\ux )=+\infty\;\hbox{ if }\;\|x_1-x_2\|_{\max}<r_0,\eqno (1.15)$$
with $r_0\in (0,r_1)$ being the diameter of the hard core. However, we
have to assume that $|U(\ux )|$ remains uniformly bounded
for $\ux$ with $r_0<|x_1-x_2\|_{\max}<r_1$.

In Section \ref{SectExtPot}, we give formal conditions upon the structure
of the potential energy terms in Hamiltonian (1.5).
A useful notion is the `shadow' $\Pi\BLam$ of a parallelepiped
$\BLam =\BLam_{L_1,L_2})(\uu )$:
$$\Pi\BLam =\Pi_1\BLam\cup\Pi_2\BLam .\eqno (1.16)$$
It can be a cube or a union of two cubes in $\R^d$,
possibly disjoint. In what
follows we call sets of this kind `cellular'.

Throughout the paper, $|\BLam |$ stands for the (Euclidean) volume
of parallelepiped $\BLam\subset\R^d\times\R^d$ and $|\Pi_j\Lam_j|$
for that of a projection cube $\Pi_j\BLam\subset\R^d$. In addition,
we use a similar notation $|{\rA}|$ for a cellular set ${\rA}\subset\R^d$.
\psn

Finally, note that an extension of the results of the present paper to the general
case with $N\geq 1$ particles is also possible. To this end, one needs to
apply the technique proposed recently in \cite{CS3}. We plan to publish such an extension in a forthcoming paper.

\section{The external potential field}
\label{SectExtPot}
\pmn

A common model of a \RFCA  {$\,$}is a Gaussian random field
on $\R^d$, with its characteristic `linearity structure';
it also serves as a `base' for producing
wider families of \RFCAs. Our presentation will take
this fact into account:
comments on properties of the external potential field
will be made from a `Gaussian perspective'. However, the
class of the external fields under consideration is much
larger and includes various `perturbations' of and `operations'
with Gaussian \RFCAs. We will not venture in this direction
in the current paper but plan to address this issue elsewhere.

A convenient way to describe our assumptions on \RFCA $\,\bfV$
is as follows. Fix a (measurable) function
$$
C:\;x,y\in\R^d\mapsto C(x,y)\in\R,
\eqno (2.1)
$$
which is (strictly) positive-definite: $\forall$ (measurable)
function $\zeta:\;\R^d\to\C$, the Lebesgue integral
$$\left\langle \zeta,\zeta\right\rangle_C:=
\int_{\R^d}\int_{\R^d}C(x,y)\zeta (x)\ovln{\zeta (y)}
{\rm d}x{\rm d}y\geq 0,\eqno (2.2)$$
and $\left\langle \zeta ,\zeta\right\rangle_C>0$ unless $\zeta = 0$ a.e.. In
these inequalities we allow the integral in the LHS
of (2.1) to equal $+\infty$. However,
given a cellular set ${\rA}\in\R^d$, we denote by
$\cL_2^C({\rA})$ the set of functions $\zeta$
with support in ${\rm A}$ and finite
$\left\langle \zeta,\zeta\right\rangle_C$:
$$\cL_2^C({\rA})=\big\{\zeta:\;\;\zeta(x)=0\;\hbox{ for }\;x\not\in{\rA}\;
\hbox{ and }\;\left\langle \zeta,\zeta\right\rangle_C<+\infty\big\},
\eqno (2.3)$$
equipped with the inner product
$$\left\langle \zeta,\eta\right\rangle_C:=
\int_{{\rA}}\int_{{\rA}}C(x,y)\zeta (x)\ovln{\eta (y)}
{\rm d}x{\rm d}y\eqno (2.4)$$
and the norm $\|\zeta\|_C:=\left\langle \zeta,\zeta\right\rangle_C^{1/2}\,$.
We will often suppose that an orthonormal basis in $\cL_2^C({\rA})$
has been given, $\{\eta^{\rA}_i,\;i=0,1,\ldots\}$, where
$$
\eta^{{\rA}}_0={\wh{\mathbf 1}}_{{\rA}}\;\hbox{ where }\;
{\wh{\mathbf 1}}_{{\rA}}=\frac{1}{Z_{{\rA}}}{\mathbf 1}_{{\rA}}.
\eqno (2.5)
$$
Here and below, ${\mathbf 1}_{{\rA}}$ stands for the indicator function
of set ${\rA}$ and the normalising constant is given by
$$Z_{{\rA}}=\left\|{\mathbf 1}_{{\rA}}\right\|_C^{1/2}.$$
In the case where $C(x,y)$ admits a bound $|C(x,y)|\leq a(x-y)$
where $\int\limits_{\R^d}a(z){\rm d}z<+\infty$, we have that
$Z_{{\rA}}\sim |{\rA}|$.
\pmn

Our first condition on \RFCA $\,\bfV$ is:
\pmn

(S) {\sl Summability}: $\forall$ cellular set ${\rA}\subset\R^d$
and $\zeta\in\cL_2^C({\rA})$:
$$\int_{{\rA}}|\zeta(y)V(y;\om )|{\rm d}y<+\infty ,\;\;\;\P-\hbox{a.s.};
\eqno (2.6)$$
consequently, we set
$$
[\zeta](\om ) =\int_{{\rA}} \zeta (y)V(y;\om ){\rm d}y,
\eqno (2.7)
$$
which yields a correctly defined (and a.s. finite) RV $[\zeta]$. In
particular, given a basis $\left\{\eta^{{\rA}}_i\right\}$ in
$\cL_2^C({\rA})$,
we obtain a sequence of RVs $\left[\eta^{{\rA}}_i\right](\om )$.
\pmn

In view of Condition (S), we can represent
a random realisation $V(x;\om)$, $x\in{\rA}$,
of external potential \RFCA $\,\bfV$ in ${\rA}$ as the sum
$$V(x;\om)=\sum_i\left[\eta^{{\rA}}_i\right](\om )
\eta^{{\rA}}_i(x),\eqno (2.8.1)$$
where the series is considered in space $\cL_2^C({\rA})$. For that
reason, we will employ an alternative notation
$\Gam^{{\rA}}_i(\om )=\left[\eta^{{\rA}}_i\right](\om )$
and the decomposition
$$V(x;\om )=\sum_i\Gam^{{\rA}}_i(\om )\eta^{{\rm A}}_i(x),\eqno (2.8.2)$$
and call RVs $\Gam^{{\rA}}_i(\om )$ coefficient random variables
in $\cL_2^C({\rA})$.
The whole sequence $\left\{\Gam^{{\rA}}_i\right\}$ of the coefficient
RVs in $\cL_2^C({\rA})$ is denoted by $\BGam_{{\rA}}$ (it is, of
course, basis-dependent). Furthermore,
if $\eta^{{\rA}}_0$ is a member of the
basis $\left\{\eta^{{\rA}}_i\right\}$ in
$\cL_2^C({\rA})$, then we denote
by $\BGam_{{\rA}}^{\bot\left[\eta_0^{{\rm A}}\right]}$ the sequence
of the remaining RVs $\left\{\Gam^{{\rA}}_i,\;i\neq 0\right\}$.

To illustrate the role of space $\cL_2^C({\rA})$, take
the example of a Gaussian \RFCA $\,\bfV$ in $\R^d$ with the covariance
kernel $C(x,y)$. In this case, RVs $[\zeta_1]$ and $[\zeta_2]$
defined in (2.7) for functions $\zeta_1,\zeta_2\in\cL_2^C({\rA})$
with $\left\langle \zeta,\zeta'\right\rangle_C=0$ are independent
normal $\cN(0,1)$.

When
${\rA}=\Pi\BLam$, with $\BLam =\BLam_{L_1,L_2}(\uu )$, the series
(2.8.1) and (2.8.2)
also determine the map
$$\begin{array}{l}\phi (\ux )\mapsto W(\ux ;\om )\phi (\ux ),\;\;
\ux=(x_1,x_2)\in\BLam ,\end{array}$$
as a (random) multiplication
operator in $\L_2(\BLam )$. This in turn
allows us to define the (random) Hamiltonian
$H_\BLam$ in
$\L_2(\BLam )$; cf. Eqn (1.4).
Given a basis in $\cL_2^C(\Pi\BLam)$,
the randomness in $H_{\BLam}$ is represented by
a sequence of coefficient RVs $\BGam_{\Pi\BLam}=
\left\{\Gam^ {\Pi\BLam}_i\right\}$.

We denote by $\mfM_{{\rA}}$ the sigma-algebra (in the space
of locally square-integrable functions $\R^d\to\R$) generated by the RVs
$[\zeta]$, $\zeta\in\cL_2^C({{\rA}})$. Then the sigma-algebra $\mfM$ is the
smallest one containing $\mfM_{{\rA}}$, $\forall$ cellular sets ${\rA}$.
The underlying probability distribution $\P$ is defined on $\mfM$;
the restriction of $\P$ to $\mfM_{\rA}$ is denoted by $\P_{\rA}$ and
the expectation relative to $\P_{{\rA}}$ by $\E_{{\rA}}$.
\pmn

Our second condition on $\bfV$ is:
\pmn

(T) {\sl Temperedness}: $\forall$ cellular set ${\rA}\subset\R^d$,
the RV
$$\oV_{{\rA}}(\om )=\hbox{sup$\,$ess}\;\Big(|V(y;\om )|:\;y\in{\rA}
\Big),\eqno (2.9)$$
is $\P_{{\rA}}$-a.s. finite, and has a finite moment
$$
\E_{{\rA}}\left(\oV_{{\rA}}\right)^{d}<+\infty.
\eqno (2.10)
$$
In this condition, we employ the representation
$$|V(x;\om)|=\left|\sum_i\Gam^{{\rA}}_i(\om )
\eta^{{\rA}}_i(x)\right|\,,\eqno (2.11)$$
following Eqn (2.8.2). Although the choice of coefficient
RVs $\Gam^{{\rA}}_i$ depends on the basis in $\cL_2^C({\rA})$,
condition T is basis-independent: when it holds for
a particular choice of the orthonotrmal basis in $\cL_2^C({\rA})$, it
also holds for all bases.

\pmn
\textbf{Remark 2.1.}\label{Rem21} Condition (2.10)
holds for a Gaussian \RFCA $\,$under a mild assumption that its samples
are `regular'. (In fact, a much stronger property takes
place, guaranteeing (2.10).) See, e.g., \cite{AW} and references therein.
\pmn

Given a pair of disjoint cellular sets ${\rA},{\rA}'\subset\R^d$,
we will be working with the conditional distribution functions
$F_{\rA}\left(\,{\ry};\;\BGam_{{\rA}\cup{\rA}'}^{\bot\left[{
\wh{\mathbf 1}_{{\rA}}}\right]}\right)$ defined as follows. Let us
fix a basis $\left\{\eta^{{\rA}\cup{\rA}'}_i,\;i=0,1,\ldots\right\}$
in $\cL_2^C({\rA}\cup{\rA}')$ with $\eta^{{\rA}\cup{\rA}'}_0
={\wh{\mathbf 1}}_{\rA}$. Then set:
$$
\begin{array}{l}
F_{\rA}\left(\,{\ry};\;\BGam_{{\rA}\cup{\rA}'}^{\bot\left[{
\wh{\mathbf 1}_{{\rA}}}\right]}\right)
:=\P\left(\left[{\wh{\mathbf 1}}_{\rA}\right]<y\Big|
\left\{\left[\eta^{{\rA}\cup{\rA}'}_j\right]:\;j\geq 1\right\}
\right),\end{array}\eqno (2.12)
$$
for the coefficient RV $\left[{\wh{\mathbf 1}}_{\rA}\right]
=\left[\eta^{\rA}_0\right]$,
given $\BGam^{\bot\left[{
\wh{\mathbf 1}_{{\rA}}}\right]}_{{\rA}\cup{\rA}'}=
\left\{\left[\eta^{\rA}_i\right]:\;i\geq 1\right\}$, a collection
of other basis-related coefficient RVs in $\cL_2^C({\rA}\cup{\rA}')$.
In fact, in Eqn (2.12) we are interested
in sets of the form ${\rA} =\Pi\BLam$
and ${\rA}'=\Pi\BLam'$ where parallelepipeds
$\BLam=\BLam_{L_1,L_2}(\uu )$ and $\BLam'=\BLam_{L_1',L_2'}(\uu')$
are distant apart, namely with
$$
\|\uu -\uu'\|_{\max} > 8\max\;\big[L_1,L_2,L'_1,L'_2\big].
\eqno (2.13)
$$
\pmn

More precisely, we exploit a simple geometric fact stated in Lemma \ref{Geom}:
\pmn

\begin{lem}\label{Geom} {\sl Consider two parallelepipeds
$\BLam=\BLam_{L_1,L_2}(\uu )$ and $\BLam'=\BLam_{L'_1,L'_2}(\uu')$
and suppose that condition} (2.13) {\sl holds true.
Then there are two possibilities (which in general do not
exclude each other):}

(i) $\BLam$ {\sl and $\BLam'$
are `completely separated', when}
$$
\dist\left[ \Pi\BLam, \Pi\BLam' \right] >0.
\eqno (2.14)
$$

(ii) $\BLam$ {\sl and $\BLam'$
are `partially separated'. In this case one (or more) of
the four possibilities can occur:}
$$\begin{array}{lll}{\rm{(A)}}
& \dist[  \Pi_1\BLam, [\Pi_2\BLam\cup \Pi\BLam' ] ] & >0,\\
{\rm{(B)}}& \dist[ \Pi_2\BLam,  [\Pi_1\BLam \cup \Pi\BLam'] ]& >0,\\
{\rm{(C)}}& \dist[ \Pi_1\BLam', [\Pi\BLam \cup \Pi_2\BLam'] ]& >0,\\
{\rm{(D)}}& \dist[ \Pi_2\BLam', [\Pi\BLam \cup \Pi_1\BLam'] ]& >0.
\end{array}
\eqno (2.15)
$$
\end{lem}
\pmn

The proof of Lemma \ref{Geom} is straightforward; it has been given in our earlier paper \cite{CS1} and applied to the lattice case, but  geometrical arguments in \cite{CS1} actually refer to parallelepipeds in Euclidean space $\R^d$.
\pmn

Pictorially,  case (ii) is where one of the cubes
$\Pi_j\BLam $, $\Pi_j\BLam'$, $j=1,2$,
is disjoint from the union of the rest of the projections of
$\BLam $ and $\BLam'$. We note that the use of the max-norm
$\|\;\|_{\max}$ is convenient here as it leads
to the constant $8$ (equal to $2$ times $4$, the number of projections
$\Pi_j\BLam$ and $\Pi_j\BLam'$, $j=1,2$) which does not depend
on dimension $d$.
\pmn

We make use of Lemma \ref{Geom} as follows. First,
for a given pair of bounded cellular sets
${\rA},{\rA}'\subset\R^d$, with ${\rA}\cap{\rA}'
=\emptyset$, we set:
$$
\begin{array}{l}\nu_{{\rA},{\rA}'}({\rb} ):=\;
{\operatornamewithlimits{\sup}\limits_{{\ry}\in \R}}\;\;
{\operatornamewithlimits{{\rm{sup\,ess}}}\limits_{
\BGam_{{\rA}\cup{\rA}'}^{\bot\left[{\wh{\mathbf 1}_{{\rA}}}\right]}}}\;\;
\Big[\;F_{{\rA}}\left(\,{\ry}+{\rb};\;\BGam_{
{\rA}\cup{\rA}'}^{\bot\left[{\wh{\mathbf 1}_{{\rA}}}\right]}\right)
-F_{{\rA}}\left(\,{\ry};\;\BGam_{{\rA}\cup{\rA}'}^{\bot\left[{
\wh{\mathbf 1}_{{\rA}}}\right]}\right)\Big]\,.
\end{array}
\eqno (2.16)
$$
Here, it is assumed that we have been given a basis
$\left\{\eta^{{\rA}\cup{\rA}'}_i,\;i=0,1,\ldots\right\}$ in
$\cL_2^C({\rA}\cup{\rA}')$, with
$\eta^{{\rA}\cup{\rA}'}_0={\wh{\mathbf 1}}_{{\rA}}$.

In particular, with ${\rA}'=\emptyset$, we have:
$$
\begin{array}{l}\nu_{\Pi\BLam ,\;\emptyset}({\rb} ):=\;
{\operatornamewithlimits{\sup}\limits_{{\ry}\in \R}}\;\;
{\operatornamewithlimits{{\rm{sup\,ess}}}\limits_{
\BGam_{{\rA}}^{\bot{\wh{\mathbf 1}_{{\rA}}}}}}\;\;
\Big[\;F_{{\rA}}\left(\,{\ry}+{\rb};\;\BGam_{
{\rA}}^{\bot\left[{\wh{\mathbf 1}_{{\rA}}}\right]}\right)
-F_{{\rA}}\left(\,{\ry};\;\BGam_{{\rA}}^{\bot{
\wh{\mathbf 1}_{{\rA}}}}\right)\Big]\,.\end{array}
\eqno (2.17)
$$
We then set:
$$
\mu_{\Pi\BLam}({\rb})=\nu_{\Pi\BLam ,\emptyset}({\rb}),
\;\;{\rb}>0.
\eqno (2.18)
$$

Next, given ${\rb}>0$, we denote:
$$
\begin{array}{l}\mu_{\BLam}^{(0)}({\rb} )
:=\;\sup\;\Big\{\nu_{\Pi\BLam,\;
\Pi\BLam'}({\rb} ):\;\;L'_1,L'_2>0,\;\uu'\in\R^d\times\R^d,\\
\qquad{}
\|\uu -\uu'\|_{\max} > 8\max\;[L_1,L_2,L'_1,L'_2]\;\hbox{ and }\;
\Pi\BLam\cap\Pi\BLam'=\emptyset\Big\}\,,
\end{array}
\eqno (2.19)
$$
$$
\begin{array}{r}\mu_{\BLam}^{(1)}({\rb} )
:=\;\sup\;\Big\{\nu_{\Pi_1\BLam,\;
\Pi_2\BLam\cup\Pi\BLam'}({\rb} ):\;\;L'_1,L'_2>0,
\qquad{}\qquad{}\\
\uu'\in\R^d\times\R^d,\;
\|\uu -\uu'\|_{\max} > 8\max\;[L_1,L_2,L'_1,L'_2]\qquad{}\\
\hbox{ and }
\Pi_1\BLam\cap\big(\Pi_2\BLam\cup \Pi\BLam'\big)=\emptyset\Big\}
\end{array}
\eqno (2.20)
$$
and
$$
\begin{array}{r}\mu_{\BLam}^{(2)}({\rb} )
:=\;\sup\;\Big\{\nu_{\Pi_2\BLam,\;
\Pi_1\BLam\cup\Pi\BLam'}({\rb} ):\;\;L'_1,L'_2>0,
\qquad{}\qquad{}\\
\uu'\in\R^d\times\R^d,\;
\|\uu -\uu'\|_{\max} > 8\max\;[L_1,L_2,L'_1,L'_2]\qquad{}\\
\text{ and } \Pi_2\BLam\cap\big(\Pi_1\BLam\cup \Pi\BLam'\big)=\emptyset\Big\}
\,.\end{array}
\eqno (2.21)
$$
(We suppose here that we have bases $
\left\{\eta^{{\Pi\BLam}\cup{\Pi\BLam'}}_i,\;i=0,1,\ldots\right\}$ in
$\cL_2^C(\Pi\BLam\cup\Pi\BLam')$, with $\eta^{{\Pi\BLam}\cup{\Pi\BLam'}}_0=
{\wh{\mathbf 1}}_{\Pi\Lam}$ in Eqn (2.19), $\eta^{{\Pi\BLam}
\cup{\Pi\BLam'}}_0={\wh{\mathbf 1}}_{\Pi_1\Lam}$ in Eqn (2.20)
and $\eta^{{\Pi\BLam}\cup{\Pi\BLam'}}_0=
{\wh{\mathbf 1}}_{\Pi_2\Lam}$ in Eqn (2.21).)
\pmn

Finally,
$$\omu_{\BLam}({\rb}):=\max\;\left[\mu_{\BLam}^{(0)}({\rb} ),
\mu_{\BLam}^{(1)}({\rb} ),\mu_{\BLam}^{(2)}({\rb} )\right].
\eqno (2.22)$$

\pmn
\textbf{Remark 2.2.}\label{Rem22} Quantity $\nu_{{\rA},{\rA}'}({\rb})$ in
Eqn (2.16) describes
a `conditioned continuity modulus' of (the
distribution function of) RV $\wh{\mathbf 1}_{{\rA}}$, given
a sample $\BGam_{{\rA}\cup{\rA}'}^{\bot{\wh{\mathbf 1}_{{\rA}}}}$.
It is important to have a grasp of the magnitude of this RV
and therefore of quantity $\nu_{{\rA},{\rA}'}({\rb})$.
As an example, we can think of
variable $\left[\wh{\mathbf 1}_{{\rA}}\right]$ as normal $\cN(0,1)$: this is
the case where \RFCA $\,{\bfV}$ is Gaussian
with covariance kernel
$C(x,y)$. In this case, $\nu_{{\rA},{\rA}'}({\rb})\leq (2\pi)^{-1/2}{\rb}$
and consequently, quantities
$\mu_{\Pi\BLam}({\rb})$ and
$\omu^{(j)}_{\Pi\BLam}({\rb})$, $j=0,1,2$, from
Eqns (2.18) and (2.19)--(2.21) satisfy a similar bound.
\psn

Concluding this section, we would like to comment on our
approach to analysis of random fields with a continuous argument,
particularly, on the meaning and use of quantities introduced in Eqns
(2.16)--(2.22). Our main intention is to provide a benefit to a
readers with a physical background.

We start by referring back to Eqn (1.3) specifying the structure of
a two-particle potential energy under consideration. In a single-particle
model, the Hamiltonian $H^{(1)}$ is simply
$-\diy{\frac{1}{2}}\Delta +V(x,\om)$.
In this case, Wegner-type bounds were proved in \cite{FHLM}, for a certain
class of external potential \RFCA{s} (including Gaussian fields).
The proof given in \cite{FHLM} is based on a special kind of
decomposition of an \RFCA in a cube $\Lam_L(u)$ (see (1.8)),
of the form
$$V(x;\om) = \varphi_0^{\Lam_L(u)}(\om) f^0_{\Lam_L(u)}(x) +
\Theta (x;\om), \; x\in\Lam_L(u),\eqno (2.23)$$
where the (scalar) random variable $\varphi_{\Lam_L(u)}(\om)$ is
independent of the `fluctuation field'
$\Theta (x;\om)$. This suggests conditioning upon values of
$\Theta (x;\om)$, so that the finite-volume Hamiltonian $H^{(1)}_{\Lam_L(u)}$
(again with Dirichlet boundary conditions) becomes a function of the
scalar random parameter $\varphi_{\Lam_L(u)}(\om)$. Naturally, its eigenvalues
are also viewed as functions of $\varphi_{\Lam_L(u)}(\om)$; their analysis,
based on the so-called
Dirichlet--Neumann bracketing and other techniques, leads to Wegner-type
bounds. These bounds are helpful for proving the existence
of the limiting density of states for single-particle model in question.
Conditions on the random field $V(x;\om)$ in \cite{FHLM} cover some
non-Gaussian \RFCA{s}, but they are somewhat restrictive even in
the Gaussian case.

In the present paper, we make use of a similar decomposition
(2.8.1)--(2.8.2). For ${\rA}=\Lam$ where $\Lam=\Pi_j\BLam (\uu)$,
$j=1,2$, it looks like
$$
V(x;\om) =\left[\eta_0^\Lam(\om)\right]{\wh\one}_\Lam(x)
+ \Xi_\Lam (x;\om), \; x\in\Lam,\;
\hbox{ where }\;\Xi_\Lam (x;\om)=\sum_{\eta_j^\Lam\in
\BGam_{{\Lam}}^{\bot{
\wh{\one}_{{\Lam}}}}}
\eqno (2.24)
$$
with a (normalised) constant `ground level' ${\wh\one}_\Lam(x)$
and a coefficient
$[\eta_0^\Lam (\om)]$ `moderately dependent' on the residual
`fluctuation field'
$\Xi (x;\om)$. Such an idea is well-known in probability theory:
in elementary courses of statistics, it is usually proved that "the
empiric mean of a Gaussian sample is independent of the empiric
variance". See, e.g., \cite{SK}.

In the context of this paper, decomposition (2.24) implies a similar
decomposition for two-particle finite-volume eigenvalues $\E^{(\BLam )}_k$
(see (1.10)). In particular, conditional on a value of $\Xi_\Lam (x;\om )$,
the eigenvalues $\E^{(\BLam )}_k$ are behaving, as functions of
$\eta_\Lam (\om )$, in a controllable fashion, which leads to our
Wegner-type bounds. In the special case when random field ${\mathbf V}$
is Gaussian, the random variables $\E^{(\BLam )}_k$ are (conditionally)
normal, with the means and variances determined by the conditions
but behaving in a controllable fashion.

\section{Wegner-type bounds}
\label{1vlbnd}

A one-volume Wegner-type bound for finite-volume Hamiltonians
is given in Theorem \ref{Thm1} below. Let
$\Sigma\left(H_{\BLam}\right)$ denote the (random) spectrum
of operator $H_{\BLam}$ from Eqn (1.4) (i.e., the collection of
its eigenvalues $E^{(\BLam}$, without their multiplicities). Next,
set
$$\begin{array}{cl}
\oW_{\BLam}(\om )&:=\max\;\Big[\;\big|W(\ux ;\om )\big|:\;
\ux\in\BLam\Big]\\ \;&\leq \oU_{\BLam}\;
+\;\oV_{\Pi_1\BLam}(\om )\;+\;\oV_{\Pi_2\BLam}(\om )\\
\;&\leq \oU_{\BLam}
+2\oV_{\Pi\BLam}(\om ). \end{array}\eqno (3.1)$$
where
$$\oU_{\BLam }={\operatornamewithlimits{\sup}\limits_{\ux
\in\BLam}}\;
|U(\ux )|,$$
and $\oV_{\Pi\BLam}$ is the supremum from Eqn (2.9).
\pmn

\begin{thm}\label{Thm1} {\sl Assume the above conditions} (I), (E0)
{\sl and} (E1) {\sl on function $U$ and external potential \RFCA
$\,\bfV$. Then, $\forall$ $E\in\R$, $L_1,L_2\geq 1$,
$\uu\in\R^d\times\R^d$ and $\eps\in (0,1)$,
with $\BLam =\BLam_{L_1,L_2}(\uu )$:
$$\begin{array}{l}\P\Big(\big[E-\eps ,E+\eps\big]\cap
\Sigma\left(H_{\BLam}\right)\neq\emptyset\Big)\\
\qquad\qquad\leq c_1\,\cdot\,\left|\BLam\right|\,\cdot\,
\E\left(E+2+\oW_{\BLam}\right)^{d}
\;\cdot\mu_{\BLam}
\left(4Z_{\Pi\BLam}\cdot\eps\right).
\end{array}\eqno (3.2)$$
Here $c_1\in (0,+\infty )$ is a constant independent of $E$,
$L_1,L_2$, $\uu$ and $\eps$.}
\end{thm}
\pmn

The expression in the RHS of (3.2) includes a
`volume' factor $\left|\BLam_{L_1,L_2}(\uu )\right|$ and a factor
$\E\left(E+2+\oW_{\BLam}\right)^{d}$
reflecting the growth of the (value of the) external potential
\RFCA $\,\bfV$.  Next, we have a factor $\mu_{\Pi\BLam}
\left(4Z_{\Pi\BLam}\cdot\eps\right)$
controlling `singularity' of the conditional distribution
functions
$$F_{\Pi\BLam}\left(\,{\ry};\;\BGam_{\Pi\BLam\cup\Pi\BLam'}^{\bot\left[{
\wh{\mathbf 1}_{\Pi\BLam}}\right]}\right)\;\hbox{and}\;
F_{\Pi_j\BLam}\left(\,{\ry};\;\BGam_{\Pi_{\oj}\BLam\cup\Pi\BLam'}^{\bot\left[{
\wh{\mathbf 1}_{\Pi_j\BLam}}\right]}\right),\;j=1,2,\;{\ry}\in\R ,$$
with a controlled distance between projections $\Pi\BLam$ and $\Pi\BLam'$.
Here $\oj=3-j$.
In a
`smooth' situation where this distribution function has a density
that is bounded uniformly in ${\ry}$ and in $L_1,L_2L'_1,L'_2$,
$\uu$, $\uu'$ satisfying (2.13), (see Remark \ref{Rem21}
above), we have that $\mu_{\BLam}\left(4Z_{\Pi\BLam}\eps\right)
\approx c'_1Z_{\Pi\BLam}\eps$ where $c'_1>0$ is a constant. Again,
such a form of the bound is important for the proof of Anderson's
localisation, the main area of application of Wegner-type
inequalities.

Our conditions on the random external potential field
are straightforward, as well as the core argument of the proof given below. However, they require from the reader
sufficient familiarity with standard constructions
used in the analysis of random fields with a continuous
argument. It may be of some help to the reader to know that
these conditions are written down by postulating, in
general terms, well-known
qualitative properties of Gaussian random fields. On the other hand, the obtained
eigenvalue concentration bounds are not optimal, and they are not supposed to be,
since our main motivation here is to lay ground for the multi-particle Multi-Scale Analysis in a Euclidean space $\R^d$, using ideas and techniques similar to those used in \cite{CS2}, \cite{CS3}.

\pmn

\begin{proof} Given a two-particle parallelepiped
$\BLam=\BLam_{L_1,L_2}(\uu )$, the randomness
in operator $H_{\BLam}(\om )$ is represented by the
sample of the \RFCA $\,\bfV_{\Pi\BLam}=\{V(x;\om ),\;x\in\Pi\BLam\}$.
More conveniently, we can refer to a sequence $\BGam_{\Pi\BLam}=
\left\{{\Gam}_i^{\Pi\BLam},\;i=0,1,\ldots\right\}$ of random
variables ${\Gam}_i^{\Pi\BLam}$ representing sample
$\bfV_{\Pi\BLam}$ in $\cL^C_2(\Pi\BLam )$:
$$V(x;\om )=\sum_{i\geq 0}{\Gam}_i^{\Pi\BLam}(\om )\eta^{\Pi\BLam}_i(x),
\;\;x\in\Pi\BLam .$$
It is convenient
to think of $H_{\BLam}(\om )$ as a family of operators
$H_{\BLam}(\Bgam )$ in $\L_2(\BLam )$
parametrised by vectors
$\Bgam =\{\gamma_i:\;i=0,1,\ldots\}$ (i.e., sample
vectors of $\BGam^{\Pi\BLam}$):
$$\begin{array}{l}\left[H_{\BLam}(\Bgam)\Bphi\right](\ux)
=\left[H^{(0)}_{\BLam}\Bphi\right](\ux)+W_\BLam(\ux;\Bgam )
\Bphi (\ux )\\
=-\,\diy{\frac{1}{2}}\sum\limits_{j=1,2}
\Delta^{(\Lam_{L_j}(u_j))}_j\Bphi(\ux)
+  \left(U(\ux)+\sum\limits_{j=1,2}\sum\limits_{i=0,1,\ldots}
{\Bgam}_i\eta^{\Pi\BLam}_i(x_j)\right) \Bphi(\ux),\\
\qquad\qquad\ux=(x_1,x_2)\in\BLam ,\;\;\Bphi\in\L_2(\BLam ).
\end{array}\eqno(3.3)
$$
Vectors $\Bgam$ are then made random, subject to
the distribution $\P_{\Pi\BLam}$ of \RFCA $\,\bfV_{\Pi\BLam}$.
Consequently, the eigenvalues $E^{(\BLam )}_k$
of $H_{\BLam}(\Bgam )$ (written in the non-decreasing order)
are parametrised by $\Bgam$,
i.e., are random variables on $\left(\R^\N,\P_{\Pi\BLam}
\right)$. We can denote them by $E^{(\BLam )}_k(\Bgam )$
but in fact, to stress the dependence on function
$W(=W(\;\cdot\;;\om))$, we will use an alternative notation
$E^{(\BLam,W)}_k$, $k=0,1,\ldots$.

In this setting, it is convenient to choose a basis
$\{\eta^{\Pi\BLam}_i,\;i=0,1,\ldots\}$ in $\cL_2^C(\Pi\BLam )$
with $\eta^{\Pi\BLam}_0={\wh{\mathbf 1}}_{\Pi\BLam}$ and work with
$\P_{\BLam}(\;\cdot\;|\Bgam^{\geq 1)})$, the
probability distribution for RV $\;\Gamma^{\Pi\BLam}_0=\left[
{\wh{\mathbf 1}}_{\Pi\BLam}\right]\;$
conditional on a given vector of values
$\Bgam^{\geq 1}=\{\gamma_j:\;j\geq 1\}$ of RVs
$\;\BGam^{\bot\left[{
\wh{\mathbf 1}_{\Pi\BLam}}\right]}_{\Pi\BLam}=
\left\{\Gam^{\Pi\BLam}_i,i\geq 1\right\}$.

Our first remark is
that the additive change $\bfv_{\Pi\BLam}\mapsto\bfv_{\Pi\BLam}
+t{\mathbf 1}_{\Pi\BLam}$ in the sample of the external
potential field ${\bfV}_{\Pi\BLam}$ generates the change
$$W(\ux )\mapsto W(\ux )+2t,\;\;\ux\in\BLam ,$$
in the value of the potential energy function $W(\ux )$,
$\ux\in\BLam$. Correspondingly, the eigenvalues
$E^{(\BLam ,W)}_0\leq E^{(\BLam ,W)}_1\leq \ldots$ and
$E^{(\BLam ,W+2t)}_0\leq E^{(\BLam ,W+2t)}_1\leq \ldots$
of the operator
$H_{\BLam}$ before and after the change are related with
$$E^{(\BLam ,W+2t)}_j=E^{(\BLam ,W)}_j+2\,t,\;\;j=0,1,\ldots .$$
Recalling that $\eta^{\Pi\BLam}_0={\mathbf 1}_{\Pi\BLam}
\big/Z_{\Pi\BLam}$,
we obtain that $\forall$ $j=0,1,\ldots$:
$$\P_{\BLam}\left(E^{(\BLam ),W}_j\in\big[E-\eps ,E+\eps\big]
\big|\Bgam^{\geq 1)}\right)\leq \mu_{\Pi\BLam}
\left(4Z_{\Pi\BLam_L}\eps\right).\eqno (3.4)$$

The second remark is that, when we pass from (3.4) to (3.2),
we only have to examine a finite number of eigenvalues
$E^{(\BLam ,W)}_j$ of Hamiltonian $H_{\BLam}$. More precisely,
denote by
$E^{(\BLam ,0)}_0\leq E^{(\BLam ,0)}_1\leq\ldots$ the eigenvalues
of the kinetic energy operator $H^0_{\BLam}$. Then the bound
$$\left|E^{(\BLam ,0)}_j-E^{(\BLam ,W)}\right|\leq
2\oV_{\Pi\BLam}\eqno (3.5)$$
implies that, $\forall$ $l=1,2,\ldots$, the conditional
probability
$$\begin{array}{l}\P\Big(\big[E-\eps ,E+\eps\big]\cap
\Sigma\left(H_{\BLam}\right)\neq\emptyset\Big|\oV_{\Pi\BLam}
\in [l-1,l)\Big)\\
\leq \#\left\{\hbox{eigenvalues of $H^0_{\BLam}$ in
$[E-\eps -l,E+\eps +l]$}\right\}\;\cdot\mu_{\Pi\BLam}
\left(4Z_{\Pi\BLam}\eps\right).\end{array}\eqno (3.6)$$

The final remark is that
$$\begin{array}{l}
\#\left\{\hbox{eigenvalues of $H^0_{\BLam}$ in $[E-\eps -l,
E+\eps +l]$}\right\}\\
\leq \#\left(\Z^d\times\Z^d\cap\left\{\ux\in\R^d
\times\R^d:\;\|\ux\|_{\rm{Euclid}}\leq E+\eps +l\right\}
\right)\\
\leq c_1\left|\BLam\right|(E+\eps +l)^{d}\end{array}\eqno (3.7)$$
where $c_1$ is the the constant from Weyl's formula.
Hence, by taking expectation, we obtain from (3.6) that
the LHS in (3.2) is
$$\leq c_1\cdot\left|\BLam\right|\cdot\E\left(E+2
+\oV_{\Pi\BLam}\right)^d\cdot\mu_{\Pi\BLam}
\left(4Z_{\Pi\BLam}\eps\right),$$
i.e., the bound (3.2) holds true. This completes the proof of
Theorem \ref{Thm1}.
\end{proof}
\pmn

We now turn to a two-volume Wegner-type bound, dealing
with a pair of
parallelepipeds $\BLam=\BLam_{L_1,L_2}(\uu )$ and
$\BLam'=\BLam_{L'_1,L'_2}(\uu')$
(more precisely, with the corresponding Hamiltonians
$H_{\BLam}$ and $H_{\BLam'}$),
under an assumption that the distance between
$\BLam$ and $\BLam'$ is of the
same order of magnitude as the size of these parallelepipeds.
Given an interval $J\subset\R$, set:
$$\begin{array}{l}
\dist\left[\Sigma\left(H_{\BLam}\right)\cap I,
\Sigma\left(H_{\BLam'}\cap J\right)\right]\\
\qquad\qquad= \inf\;\left[\;
\left|E^{(\BLam )}_k-E^{(\BLam')}_{k'}\right|:\;\;
E^{(\BLam )}_k,E^{(\BLam')}_{k'}\in J\;
\right],\end{array}\eqno (3.8)$$
where $E^{(\BLam )}_k$ are the eigenvalues of $H_{\BLam}$
and $E^{(\BLam')}_{k'}$ those of $H_{\BLam'}$,\\ $k,k'=0,1,\ldots$.
\pmn

Because the potential
$$W(\ux)=U(\ux )+g[V(x_1;\om )+V(x_2;\om)]\eqno (3.9)$$
is a symmetric
function of the pair $\ux=(x_1,x_2)\in\R^d\times\R^d$, with
$$W(\cS\ux)=W(\ux),\;\hbox{
where $\cS:(x_1,x_2)\mapsto (x_2,x_1)$.}$$
Consequently, the spectra $\Sigma \left(H_\BLam\right)$ and
$\Sigma\left(H_{\cS(\BLam )}\right)$ are identical, and the
same is true for $\Sigma \left(H_{\BLam'}\right)$ and
$\Sigma\left(H_{\cS(\BLam')}\right)$.
\pmn

\begin{thm} \label{Thm2} {\sl $\forall$ $L_1,L_2,L'_1,L'_2>1$,
$\uu,\uu'\in\R^d\times\R^d$ with
$$
\min\, \{\|\uu -\uu'\|_{\max}, \|\cS(\uu) -\uu'\|_{\max} \}
 > 8\max [L_1,L_2,L'_1,L'_2]
 \eqno(3.10)
$$
and $\forall$ $\eps\in(0,1)$ and interval
$J=[{\rb}-\delta,{\rb}+\delta]\subset\R$, where ${\rb}\in\R$ and $\delta >0$:}
$$\begin{array}{l}
\P\Big(\dist\left[\Sigma\left(H_{\BLam}\right)\cap J,
\Sigma\left(H_{\BLam'}\right)\cap J\right]
\leq \eps \Big)\qquad{}\qquad{}\\
\qquad\leq\;c_2\,\cdot\,
\left|\BLam\right|\;\left|\BLam'\right|\;
\E\Big[({\rb}+\delta +1+\oW_{\BLam})^d({\rb}+\delta +1+\oW_{\BLam'})^d\Big]\\
\qquad\qquad\qquad\qquad\times\;
{\operatornamewithlimits{\max}\limits_{j=1,2}}\;
\left[\omu_{\Pi\BLam}\left(4\eps\cdot Z_{\Pi\BLam}\right),\;
\omu_{\Pi\BLam'}\left(4\eps\cdot Z_{\Pi\BLam'}\right)\right]\,
 .\end{array}\eqno (3.11)$$
\end{thm}

\pmn
\textbf{Remark 3.1.}\label{Rem31} As in the case of the one-volume Wegner-type
estimate (3.2), the RHS in (3.11) is composed by `volume'
factors $\left|\BLam\right|$ and $\left|\BLam'\right|$ and
factors
$$\E\Big[({\rb}+\delta +1+\oW_{\BLam})^d({\rb}+\delta +1
+\oW_{\BLam'})^d\Big]\eqno (3.12)$$
and
$${\operatornamewithlimits{\max}\limits_{j=1,2}}\;
\left[\omu_{\Pi\BLam}\left(4\eps Z_{\Pi\BLam}\right),\;
\omu_{\Pi\BLam'}\left(4\eps Z_{\Pi\BLam'}\right)\right]\,.\eqno (3.13)$$
These factors assess various aspects of
randomness introduced in Hamiltonians $H_{\BLam}$ and $H_{\BLam'}$.
For us, an immediate use of Theorems \ref{Thm1} and \ref{Thm2} is in
proving Anderson's localisation
in a two-particle model (with interval $J=[{\rb}-\delta,{\rb}+\delta]$
at the `edge of the spectrum, specified by further
assumptions on RF $\bfV$).
\pmn

\begin{proof} Owing to Lemma \ref{Geom}, parallelepipeds  $\BLam $ and
$\BLam'$ obeying (3.10) satisfy either (i) or (ii), i.e. they are
either completely
or partially separated.  Consider first case (i) (complete separation).
Write
$$\begin{array}{l}
\P\Big(\dist\left[\Sigma\left(H_{\BLam}\right)\cap J,
\Sigma\left(H_{\BLam'}\right)\cap J\right]\leq \eps \Big)\\
\qquad= \esm{ \P\Big(\dist\left[\Sigma\left(H_{\BLam}\right)
\cap J,
\Sigma\left(H_{\BLam'}\right)\cap J\right] \leq \eps \, \Big|
\,\bfV_{\Pi\BLam'} \Big)}.
\end{array}\eqno (3.14)$$
Note first that, under
conditioning  in Eqn (3.14), the eigen-values
$E^{(\BLam')}_{k'}$,
$k'=0,1,\ldots$, forming the set
$\Sigma\left(H_{\BLam'}\right)$ are non-random. Therefore, it makes
sense to use the following inequality:
$$\begin{array}{l}
 \P\Big(\dist\left[\Sigma\left(H_{\BLam}\right),
\Sigma\left(H_{\BLam'}\right)\right] \leq \eps \, \Big|
\,\bfV_{\Gamma'} \Big)\\
\qquad{}\leq  N_{\BLam'}\left(J;\oW_{\BLam'}\right)
{\operatornamewithlimits \sup\limits_{E\in J}}\;
\P\Big(\dist\left[\Sigma\left(H_{\BLam}\right)\cap J,
E\right ] \leq \eps\Big|\bfV_{\Pi\BLam'}\Big),
\end{array}\eqno (3.15)$$
where $ N_{\BLam'}\left(J;\oW_{\BLam'}\right)$ is
the number of the eigenvalues $E^{(\BLam')}_{k'}$ that
can `eventuallly' fall in $J$. As in (3.11), for $J=[a,b]$ and
with $E^0_{k'}$ standing for the eigenvalues of $H^0_{\BLam'}$, we have:
$$\begin{array}{l}N_{\BLam'}(J;\oW_{\BLam'})\\
\quad\leq \#\left\{\hbox{eigenvalues $E^0_{k'}\in
\big[\big(a-\oW_{\BLam'}\big)^+,b+\oW_{\BLam'}\big]$}\right\}\\
\quad\leq b_{2d}(L'_1L'_2)^{d}(b+\oW_{\BLam'})^{d}.
\end{array}\eqno (3.16)$$

Next, as in Theorem \ref{Thm1},
$$\begin{array}{l}
\P\Big(\dist\left[\Sigma\left(H_{\BLam}\right), E\right]
\leq \eps\Big|\BGam_{\Pi\BLam'}\Big)\\
\qquad\leq c_1\,\cdot\,\left|\BLam\right|
\E\Big[\left(E+2+\oW_{\BLam}\right)^{d}\Big|
\BGam_{\Pi\BLam'}\Big]\quad{}\\ \;\\
\qquad\quad\times
\omu_{\Pi\BLam}\left(4\eps Z_{\Pi\BLam}\right).
\end{array}\eqno (3.17)$$
This yields
$$\begin{array}{l}
 \P\Big(\dist\left[\Sigma\left(H_{\BLam}\right),
\Sigma\left(H_{\BLam'}\right)\right] \leq \eps \, \Big|
\,\BGam_{\Pi\BLam'}\Big)\qquad\qquad\qquad{}\\
\qquad\leq c^{(1)}_2\;|\BLam|\;|\BLam'|\;
(b+2+\oW_{\BLam'})^d\;\E\big[(b+2+\oW_{\BLam})^d
\big|\BGam_{\Pi\BLam'}\big]\qquad{}\\
\qquad\quad\times
\omu_{\Pi\BLam}\left(4\eps Z_{\Pi\BLam}\right),
\end{array}\eqno (3.18)$$
implying that
$$\begin{array}{l}
\P\Big(\dist\left[\Sigma\left(H_{\BLam}\right),
\Sigma\left(H_{\BLam'}\right)\right]\leq \eps \Big)\\
\qquad\leq c^{(1)}_2\;|\BLam|\;|\BLam'|\;\E\big[(b+2
+\oW_{\BLam})^d(b+2+\oW_{\BLam'})^d\big]\qquad{}\\
\qquad\quad\times
\omu_{\Pi\BLam}\left(4\eps Z_{\Pi\BLam}\right),
\end{array}\eqno (3.19)$$
where $c^{(1)}_2\in (0,+\infty )$ is a constant.
\pmn

(ii) Now consider the case of partial separation.
For example, assume case (A) from Lemma \ref{Geom} (see Eqn (3.11)),
when projection
$\Pi_1\BLam$ is disjoint from the union of the rest of the
projections of $\BLam$ and $\BLam'$:
$$\Pi_1\BLam\cap\left[\Pi_2\BLam\cup\Pi\BLam')\right]
=\emptyset.\eqno (3.20)$$
We then estimate the probability in the LHS of (3.16)
with the help of the conditional expectation
$$\begin{array}{l}
\P\Big(\dist\left[\Sigma\left(H_{\BLam}\right),
\Sigma\left(H_{\BLam'}\right)\right]\leq \eps \Big)\\
\qquad= \esm{ \P\Big(\dist\left[\Sigma\left(H_{\BLam}\right),
\Sigma\left(H_{\BLam'}\right)\right] \leq \eps \, \Big|
\,\BGam_{\Pi_2\BLam\cup\Pi\BLam'} \Big)\Big|\BGam_{\Pi\BLam'}}.
\end{array} \eqno(3.21)$$

Note that, owing to (3.20), the sigma-algebra generated by
$\BGam_{\Pi_2\BLam\cup\Pi\BLam'}$ does not
include any of the RVs $\Gam^{\Pi_1\BLam}_j$ forming $\BGam_{\Pi_1\BLam}$.
Thus, the argument used in the proof of Theorem \ref{Thm1} is
still applicable if, instead of $\P$, we work with the
probability distribution $\P_{\Pi_1 \BLam}\big(\;\cdot\;\big|
\BGam^{\geq 1}_{\Pi_1\BLam},\BGam_{\Pi_2\BLam\cup\Pi\BLam'}\big)$,
conditional on
$\bfV_{\Gamma_2\cup\Gamma'}\big)$ and restricted to the
the sigma-algebra generated by $\Gam_{{\mathbf 1}_{\Pi_1\BLam}}$.

This allows us to write
$$\begin{array}{l}\P\Big(\dist\left[\Sigma\left(H_{\BLam}\right),
\Sigma\left(H_{\BLam'}\right)\right] \leq \eps \, \Big|
\,\BGam^{\geq 1}_{\Pi_1\BLam},\BGam_{\Pi_2\BLam\cup\Pi\BLam'} \Big)\\
\qquad\leq c^{(2)}_2\;|\BLam|\;|\BLam'|\;
(b+2+\oW_{\BLam'})^d\\
\qquad\quad\times\E\big[(b+2+\oW_{\BLam})^d
\big|\BGam^{\geq 1}_{\Pi_1\BLam},
\BGam_{\Pi_2\BLam\cup\Pi\BLam'}\big]\;
\omu_{\Pi\BLam}\left(4\eps Z_{\Pi\BLam}\right)
\end{array}\eqno (3.22)$$
and
$$\begin{array}{l}\P\Big(\dist\left[\Sigma\left(H_{\BLam}\right),
\Sigma\left(H_{\BLam'}\right)\right] \leq \eps \,\Big)\\
\qquad\leq c^{(2)}_2\;|\BLam|\;|\BLam'|\;
\E\Big[(b+2+\oW_{\BLam})^d(b+2+\oW_{\BLam'})^d\Big]
\qquad{}\\
\qquad\quad\times
\omu_{\Pi\BLam}\left(4\eps Z_{\Pi\BLam}\right).
\end{array}\eqno (3.23)$$
We are then in position to
deduce the required bound for the the conditional probability
in the LHS of (3.21).

If, instead of (3.20), we have one of the other disjointedness
relations (B)-(D) in Eqn
(3.11), then the argument is conducted in a similar fashion.
Specifically, in case (B) we
swap projections $\Pi_1\BLam$ and $\Pi_2 \BLam$ in the above argument.
Furthermore, in cases (C) and (D), we should exchange $\uu$ and $\uu'$ as
compared to arguments in cases (A) and (B).
\end{proof}
\pmn

We now briefly comment on the case where the interaction potential
has a hard core (cf. (1.15)). In this case Hamiltonians $H$ and
$H_\BLam$ in Eqns (1.1) and (1.5) act in subspaces
$$\L_2\left(\big(\R^d\times\R^d\big)\setminus\D_{r_0}\right)=
\Big\{\Bphi\in\L_2(\R^d\times\R^d):\Bphi (\ux )=0\;\hbox{for}\;
\ux\in\D_{r_0}\Big\}\eqno (3.24)$$
and
$$\L_2\left(\BLam\setminus\D_{r_0}\right)=
\Big\{\Bphi\in\L_2(\R^d\times\R^d):\Bphi (\ux )=0\;\hbox{for}\;
\ux\in\D_{r_0}\Big\}\eqno (3.25)$$
where
$$\D_{r_0}=\big\{\ux =(x_1,x_2)\in\R^d\times\R^d:\;
\|x_1-x_2\|_{\max}<r_0\big\}.
\eqno (3.26)$$
More precisely, $H$ and
$H_\BLam$ carry additional Dirihclet's boundary conditions
on $\partial\D_{r_0}$. However, the scheme of the proof of
Theorems \ref{Thm1} and \ref{Thm2} remains unchanged, and all arguments carry
through.
\psn


\subsection*{Acknowledgment}
We thank The Isaac Newton Institute, University of
Cambridge, for hospitality during the programme
``Mathematics and Physics of the Anderson Localization:
50 years after'' (July--December, 2008).
YS thanks IHES, Bures-sur-Yvette, for hospitality during visit in 2008.
\end{document}